# OPTIMIZING CHANNEL SELECTION FOR SEIZURE DETECTION

*V. Shah, M. Golmohammadi, S. Ziyabari, E. Von Weltin, I. Obeid and J. Picone*

Neural Engineering Data Consortium, Temple University, Philadelphia, Pennsylvania, USA
{vinitshah, meysam, saeedeh, eva.vonweltin, obeid, picone}@temple.edu

**Abstract—** Interpretation of electroencephalogram (EEG) signals can be complicated by obfuscating artifacts. Artifact detection plays an important role in the observation and analysis of EEG signals. Spatial information contained in the placement of the electrodes can be exploited to accurately detect artifacts. However, when fewer electrodes are used, less spatial information is available, making it harder to detect artifacts. In this study, we investigate the performance of a deep learning algorithm, CNN-LSTM, on several channel configurations. Each configuration was designed to minimize the amount of spatial information lost compared to a standard 22-channel EEG. Systems using a reduced number of channels ranging from 8 to 20 achieved sensitivities between 33% and 37% with false alarms in the range of [38, 50] per 24 hours. False alarms increased dramatically (e.g., over 300 per 24 hours) when the number of channels was further reduced. Baseline performance of a system that used all 22 channels was 39% sensitivity with 23 false alarms. Since the 22-channel system was the only system that included referential channels, the rapid increase in the false alarm rate as the number of channels was reduced underscores the importance of retaining referential channels for artifact reduction. This cautionary result is important because one of the biggest differences between various types of EEGs administered is the type of referential channel used.

## I. Introduction

An electroencephalogram (EEG) is a very popular non-invasive tool for recording signals and diagnosing brain-related illnesses [1]. The 10-20 electrode configuration is by far the most popular standard worldwide for conducting EEG tests [2] and provides clinicians an adequate amount of information about the signal to make a diagnosis. Higher density EEGs are popular in research communities for their superior localization capabilities, but are still not common in clinical practice. Though the increased density of the electrode grid does provide additional information, this information is not significantly more informative and does not justify the additional disk space required to archive the data.

The TUH EEG Corpus (TUEEG) [3], which is the subject of this study, is the world's largest publicly accessible archive of clinical EEG recordings. It contains over 40 unique channel configurations. Many of these configurations were created to assist in the diagnosis of specific diseases. The most striking difference in these configurations is the manner in which ground and reference is used when a differential montage is constructed [4][5]. Since EEG signals are very low in voltage and quite noisy, grounding and/or referencing plays a key role in one's ability to collect clean signals.

In this paper, we focus on an important subset of TUEEG known as the TUH EEG Seizure Corpus (TUSZ) [6]. Over 90% of these files use the 19-channel configuration shown in Figure 1 [7]. We have applied a combination of longitudinal and transverse bipolar montages, referred to as a TCP montage [7], to create 22 channel differential-bindings with a focus on focal regions of the scalp. This montage is also summarized in Figure 1.

TUSZ has been manually annotated for diverse morphologies of seizure events. We have introduced a deep learning architecture [9] that achieves a very low false positive rate (FPR). This system integrates convolutional neural networks (CNNs) with recurrent neural networks (RNNs) to deliver state of the art performance. This doubly deep recurrent convolutional structure models both spatial relationships (e.g., cross-channel dependencies) and temporal dynamics (e.g., events such as spikes).

The integration of CNNs and long-short term memory (LSTM) units does a much better job rejecting artifacts. Artifacts and events such as wicket spikes, rectus muscle and electrode-pop artifacts are easily confused with spike and wave discharges because they often appear on only a few channels similar to the way seizure events present themselves. The depth of the convolutional network is important since the top convolutional layers tend to learn generic features while the deeper layers learn dataset specific features. The convolutional LSTM architecture with proper initialization and regularization delivers 30% sensitivity at 6 false alarms per 24 hours [10].

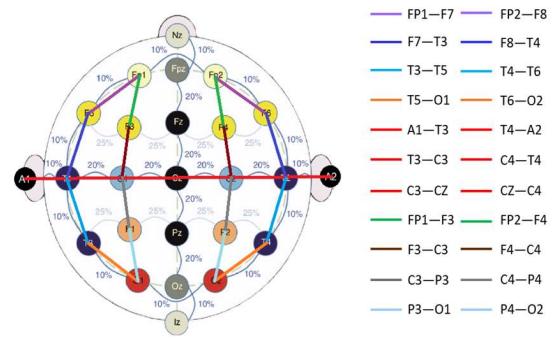

Figure 1. Electrode locations for a standard 10-20 system with a defined 22-channel TCP montage.

Feature extraction typically relies on time frequency representations of the signal. Though we can replace traditional model-based feature extraction with deep learning-based approaches that operate directly on the sampled data, in this work we focus on the use of traditional cepstral-based features. In our current system, we use a traditional linear frequency cepstral coefficient-based feature extraction approach (LFCCs) [5][8]. We also use first and second derivatives of the features since these improve performance. Though we can replace traditional model-based feature extraction with deep learning-based approaches that operate directly on the sampled data, or more advanced discriminative features, these have not yet produced substantial improvements in performance for this application.

Neurologists typically review EEGs in 10 sec windows and identify events with a temporal resolution of approximately 1 sec. Following this approach, we chose to analyze the signal in 1 sec epochs, and further divide this interval into 10 frames of 0.1 secs each so that features are computed every 0.1 seconds (referred to as the frame duration) using 0.2 second analysis windows (referred to as the window duration). The output of our feature extraction process is a feature vector of dimension 26 for each of the 22 channels, with a frame duration of 0.1 secs. This optimized system produces 39% sensitivity and 90% specificity with 23 false alarms (FA) per 24 hours [9]. This will be our baseline system.

Our focus in this study is to optimize the selection of channels. This serves two purposes. First, it reduces the dimensionality of the problem. Second, and more importantly, our goal is to find a minimal number of channels that are common across all EEGs that can provide reasonable levels of performance. Otherwise, the system will have to adapt to the unique channel configuration of each EEG or clinical site, and this is an extremely complex process.

The results presented in this paper use the Any Overlap scoring method [11] in which true positives are counted when the hypothesis overlaps with one or more reference annotations. False positives correspond to events in which the hypothesis annotations do not overlap with any of the reference annotations. This method of scoring is popular in the EEG research community. The relative rankings of the systems are not sensitive to the scoring method, though the absolute numbers do change slightly.

## II. CHANNEL SELECTION

EEGs are used to diagnose a wide variety of pathologies. Applications include obvious things like seizure detection and prediction. But an EEG today is also being used to diagnose psychological disorders, sleep disorders and head injuries. Further, an EEG is used to monitor the impact of drug interventions. For each specific task, spatial information plays a major role. For example, electrodes placed near the occipital lobe capture brain activity related to vision whereas mid-parietal region electrodes collect information related to waking consciousness.

In this study, we have focused on seizure detection. We emphasize the importance of using domain knowledge in the selection of channel configurations instead of using an ad hoc selection process. An overview of the channel selection process is given in Figure 2. When reducing the number of channels from 22 to 20, we removed the referential channels A1 and A2. These are attached to the patient's ears and are generally very susceptible to noise. Additionally, all brain events occurring on those channels can also be observed on electrodes T3 and T4.

Frontal Polar (FP1 & FP2) channels are mostly ignored because only 36% of frontal seizures can be observed on scalp EEGs making automatic detection of frontal lobe seizures very difficult [12]. The CZ electrode is utilized throughout all configurations because, due to its location at the center of the scalp and because it is attached to 6 adjacent electrodes in the TCP montage, the CZ electrode is able to detect seizures occurring in both hemispheres better than any other single electrode. Only one of the occipital (O1 & O2) electrodes have been considered in 4 and 2 channel configurations because the occipital electrodes are always placed close to each other. Consequently, it is likely that seizure events occurring near one of the occipital electrodes will appear on the other as well.

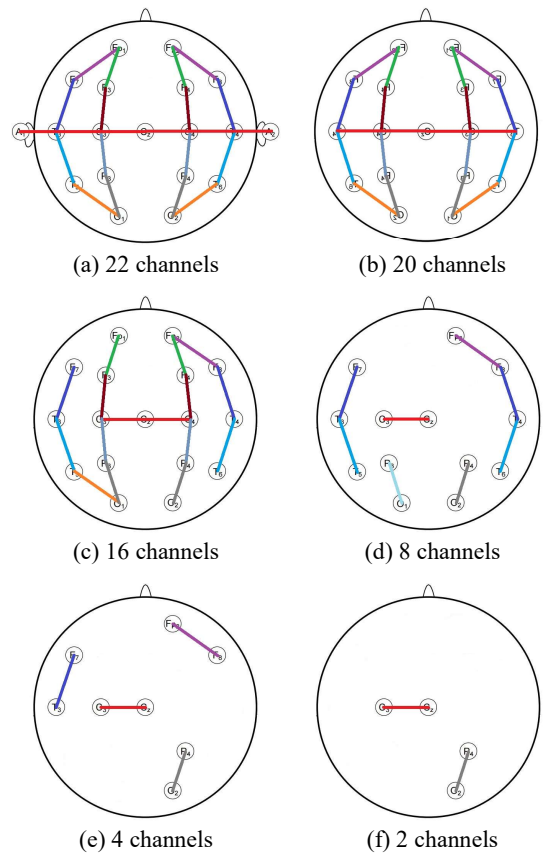

(a) 22 channels  (b) 20 channels

(c) 16 channels  (d) 8 channels

(e) 4 channels  (f) 2 channels

Figure 2. An overview of the channel selection strategies that were employed to reduce the number of channels.

### III. EXPERIMENTAL DESIGN AND ANALYSIS

For this study, we have used a baseline system that integrates CNNs and LSTMS, as shown in Figure 3. The input tensors are fed to a CNN stage that typically consists of 3 layers of 2D CNN layers with 16 kernels of size of 3×3 and max-pooling layers of size 2×2 to effectively reduce the dimensionality of the input. Dropout layers are added at the end of each layer except the very last one to avoid overfitting. The output is then flattened and fed to a 1D CNN network which acts as a fully connected network. The output of this pass is fed to a bidirectional LSTM stage. Exponential Linear Units (ELU) are used as the activation functions for all stages except the last stage, which uses a sigmoid activation function. A mean-square error loss function and Adam optimizer are also used. Postprocessing is used on the system output to reduce the false alarm rate.

In Table 1 we summarize the results for each of the channel configurations shown in Figure 2. The system with the 22-channel configuration, as expected, outperforms the other systems. The 20-channel, 16-channel and 8-channel configurations produce moderate reductions in performance. The 4-channel and 2-channel configurations perform poorly because these configurations lack spatial context.

Unfortunately, the typical system defined here cannot be applied identically for all the channel configurations that we have defined for this study because dimensionality reduction on a small number of channels is a problem. Applying max-pooling with a 2×2 matrix on all the layers when using 2, 4, and 8, channels is not possible. To make a fair comparison and to understand the behavior of a system on low dimensional tensors we have used two separate approaches for low-dimensional channel configurations. First, we simply keep the dimensionality of channel tensor intact. Second, we remove one or more CNN layers whenever we face dimensionality reduction issues. Modification in number of CNN layers can be observed in the second column in Table 1.

An ROC curve, which depicts the true positive rate (TPR) vs. the false positive rate (FPR), is shown in Figure 4. We compare four systems: 22, 16, 8 and 4 channels. The 22-channel system clearly outperforms the other three reduced-channel configurations. The performance differences are greatest for low values of FPR, which is the region of most interest in this application. On the other hand, when the FPR is high, the performance

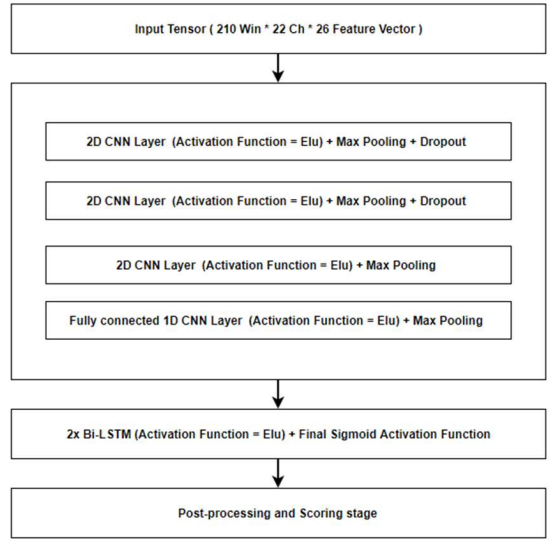

Figure 3. A block diagram of the baseline system.

between these systems is minimal.

We also observe that the performance differences between 16-channel and 8-channel configurations are small with the 8-channel system performing slightly better when the FPR is low. This seems to validate the process used to select these channel configurations that was based on significant amounts of domain knowledge.

Table 1. Performance vs. channel configuration

| Ch. | 2D CNN Layers | Sensitivity (%) | Specificity (%) | FA/24 Hours |
|---|---|---|---|---|
| 22 | 3 | 39.15 | 90.37 | 22.83 |
| 20 | 3 | 34.54 | 82.07 | 49.25 |
| 16 | 3 | 36.54 | 80.48 | 53.99 |
| 8 | 3 | 33.44 | 85.51 | 38.19 |
| 4 | 3 | 33.11 | 39.32 | 325.54 |
| 8 | 2 | 30.66 | 88.79 | 28.57 |
| 4 | 1 | 34.09 | 39.00 | 332.15 |
| 2 | 3 | 31.15 | 40.82 | 308.74 |

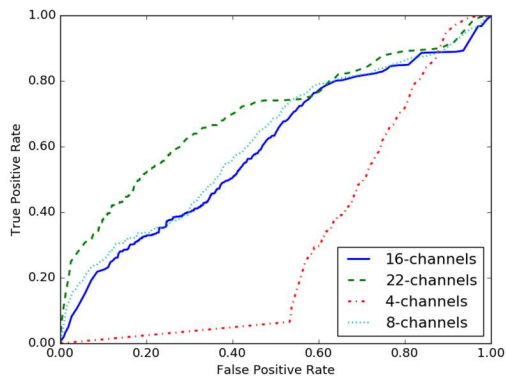

Figure 4. ROC curves for 22, 16, 8 and 4 channels

Next, we conducted an experiment to investigate the importance of including the referential channels A1 and A2, referred to collectively as Ax. Table 2 presents a comparison the 2, 4, 8 and 16 channel configurations to the same configurations with Ax added. We also provide an ROC curve in Figure 5. The ROC curves demonstrate that gap in performance between the 18-channel system and the 10-channel system is much greater than that achieved without the additional channels. Further, overall performance with $A_x$ is better than without.

To further probe this, in Figure 6, we compare an 18-channel configuration with $A_x$ channels to a 16-channel configuration without the $A_x$ channels. The system using referential channels performs better at low FPR than the system without referential channels, and this improvement in performance is not simply due to the increased channel count. Instead it is an indication that the referential channels are providing meaningful information, especially at low FPRs.

## IV. SUMMARY

In this paper, we have investigated the impact of referential channels on seizure detection performance. We have explored this using a framework based on a hybrid CNN-LSTM deep learning system. Not surprisingly, using all channels from a 10-20 EEG configuration gave best performance: 39.15% sensitivity and 90.37% specificity with 22.83 FA per 24 hours. Selection a moderately reduced number of channels (e.g.,

Table 2. A comparison of performance demonstrating the impact of including the $A_x$ channels

| No. Chan. | | Sensitivity (%) | | FA/24 Hours | |
|---|---|---|---|---|---|
| w/ $A_x$ | w/o $A_x$ | w/ $A_x$ | w/o $A_x$ | w/ $A_x$ | w/o $A_x$ |
| 22 | 20 | 39.15 | 34.54 | 22.83 | 49.25 |
| 18 | 16 | 36.65 | 36.54 | 37.33 | 53.99 |
| 10 | 8 | 30.94 | 33.44 | 283.18 | 38.19 |
| 6 | 4 | 34.36 | 34.09 | 58.15 | 332.15 |
| 4 | 2 | 33.06 | 31.15 | 47.53 | 308.74 |

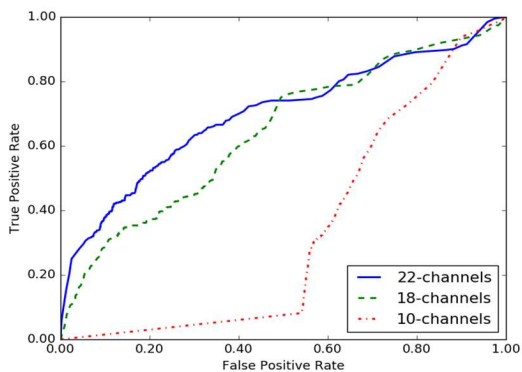

Figure 5. ROC curves for 22, 18 and 10 channel configurations that include the $A_x$ channels

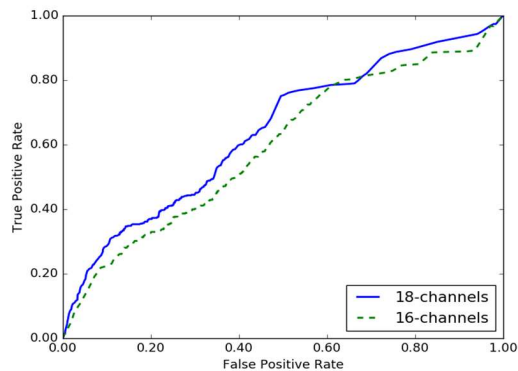

Figure 6. 18-channels w/ $A_x$ vs. 16-channels w/o $A_x$

16 and 8) resulted in a small but measurable degradation in performance. Adding referential channels to these configurations improved performance particularly in the low FPR region of primary interest in this application.

Deep learning systems are extremely sensitive to training conditions. Initialization of models and randomization of the data play a far too significant role in the overall performance. This complicates these types of parameter studies because the system must be individually optimized for each condition. This is an ongoing issue that we are addressing in future research.

Also, we demonstrated that network architectures needed to change for the low-order systems. For example, the 4-channel system in Table 1 used only one 2D CNN layer. These types of optimizations are another reason these parametric studies must be carefully designed.

Finally, since the use of referential channels varies significantly across EEG type, clinical site, neurologist, etc. Better techniques to reduce the sensitivity of performance to these referential channels is needed.

## ACKNOWLEDGMENTS

Research reported in this publication was most recently supported by the National Human Genome Research Institute of the National Institutes of Health under award number U01HG008468. The content is solely the responsibility of the authors and does not necessarily represent the official views of the National Institutes of Health. This material is also based in part upon work supported by the National Science Foundation under Grant No. IIP-1622765. Any opinions, findings, and conclusions or recommendations expressed in this material are those of the author(s) and do not necessarily reflect the views of the National Science Foundation.

## REFERENCES

[1] T. Yamada and E. Meng, *Practical guide for clinical neurophysiologic testing: EEG*. Philadelphia, Pennsylvania, USA: Lippincott Williams & Wilkins, 2009.

[2] W. Tatum, A. Husain, S. Benbadis, and P. Kaplan, *Handbook of EEG Interpretation*. New York City, New York, USA: Demos Medical Publishing, 2007.

[3] I. Obeid and J. Picone, "The Temple University Hospital EEG Data Corpus," *Front. Neurosci. Sect. Neural Technol.*, vol. 10, p. 196, 2016.

[4] I. Obeid and J. Picone, "Machine Learning Approaches to Automatic Interpretation of EEGs," in *Biomedical Signal Processing in Big Data*, 1st ed., E. Sejdik and T. Falk, Eds. Boca Raton, Florida, USA: CRC Press, 2017 (in press).

[5] S. Lopez, M. Golammadi, I. Obeid, and J. Picone, "An Analysis of Two Common Reference Points for EEGs," *Proceedings of the IEEE Signal Processing in Medicine and Biology Symposium*, 2016, pp. 1–4.

[6] M. Golmohammadi, V. Shah, S. Lopez, S. Ziyabari, S. Yang, J. Camaratta, I. Obeid, and J. Picone, "The TUH EEG Seizure Corpus," *Proceedings of the American Clinical Neurophysiology Society Annual Meeting*, 2017, p. 1.

[7] ACNS, "Guideline 6: A Proposal for Standard Montages to Be Used in Clinical EEG," Milwaukee, WS, USA, 2006 (available at *http://www.acns.org/pdf/guidelines/Guideline-6.pdf*).

[8] A. Harati, M. Golmohammadi, S. Lopez, I. Obeid, and J. Picone, "Improved EEG Event Classification Using Differential Energy," *Proceedings of the IEEE Signal Processing in Medicine and Biology Symposium*, 2015, pp. 1–4.

[9] M. Golmohammadi, S. Ziyabari, V. Shah, I. Obeid, and J. Picone, "Deep Architectures for Automated Seizure Detection in Scalp EEGs," submitted to the AAAI Conference on Artifical Intelligence, 2018, pp. 1–8 (available at: *https://www.isip.piconepress.com/publications/unpublished/conferences/2018/aaai/deep_learning/*).

[10] Golmohammadi, M., Ziyabari, S., Shah, V., Obeid, I., & Picone, J. (2017). Gated Recurrent Networks for Seizure Detection. Submitted to the IEEE Signal Processing in Medicine and Biology Symposium (pp. 1–5) (available at *https://www.isip.piconepress.com/publications/unpublished/conferences/2017/ieee_spmb/rnn/*).

[11] Ziyabari, S., Shah, V., Golmohammadi, M., Obeid, I., & Picone, J. (2017). An Analysis of Objective Performance Metrics for Automatic Interpretation of EEG Signal Events. Submitted to the *Journal of Clinical Neurophysiology* (available at: *https://www.isip.piconepress.com/publications/unpublished/journals/2017/jcn/metrics/*).

[12] W. Deburchgraeve et al., "Automated neonatal seizure detection mimicking a human observer reading EEG," *Clin. Neurophysiol.*, vol. 119, no. 11, pp. 2447–2454, 2008.